\documentclass[twocolumn,showpacs,preprintnumbers,amsmath,amssymb,superscriptaddress]{revtex4-1}

\usepackage{graphicx}
\usepackage{graphics}
\usepackage{dcolumn}
\usepackage{bm}
\usepackage{color}

\begin{document}

\title{Revisiting Feynman's ratchet with thermoelectric transport theory}

\author{Y. Apertet}\email{yann.apertet@gmail.com}
\affiliation{Institut d'Electronique Fondamentale, Universit\'e Paris-Sud, CNRS, UMR 8622, F-91405 Orsay, France}
\affiliation{Lyc\'ee Jacques Pr\'evert, F-27500 Pont-Audemer, France}
\author{H. Ouerdane}
\affiliation{Russian Quantum Center, 100 Novaya Street, Skolkovo, Moscow region 143025, Russia}
\affiliation{Laboratoire Interdisciplinaire des Energies de Demain (LIED), UMR 8236 Universit\'e Paris Diderot, CNRS, 5 Rue Thomas Mann, 75013 Paris, France}
\author{C. Goupil}
\affiliation{Laboratoire Interdisciplinaire des Energies de Demain (LIED), UMR 8236 Universit\'e Paris Diderot, CNRS, 5 Rue Thomas Mann, 75013 Paris, France}
\author{Ph. Lecoeur}
\affiliation{Institut d'Electronique Fondamentale, Universit\'e Paris-Sud, CNRS, UMR 8622, F-91405 Orsay, France}

\date{\today}

\begin{abstract}
We show how the formalism used for thermoelectric transport may be adapted to Smoluchowski's seminal thought experiment also known as Feynman's ratchet and pawl system. Our analysis rests on the notion of \emph{useful flux}, which for a thermoelectric system is the electrical current and for Feynman's ratchet is the effective jump frequency. Our approach yields original insight into the derivation and analysis of the system's properties. In particular we define an \emph{entropy per tooth} in analogy with the \emph{entropy per carrier} or Seebeck coefficient, and we derive the analog to Kelvin's second relation for Feynman's ratchet. Owing to the formal similarity between the heat fluxes balance equations for a thermoelectric generator (TEG) and those for Feynman's ratchet, we introduce a distribution parameter $\gamma$ that quantifies the amount of heat that flows through the cold and hot sides of both heat engines. While it is well established that $\gamma = 1/2$ for a TEG, it is equal to 1 for Feynman's ratchet. This implies that no heat may be rejected in the cold reservoir for the latter case. Further, the analysis of the efficiency at maximum power shows that the so-called Feynman efficiency corresponds to that of an exoreversible engine, with $\gamma=1$. Then, turning to the nonlinear regime, we generalize the approach based on the convection picture and introduce two different types of resistance to distinguish the dynamical behavior of the considered system from its ability to dissipate energy. We finally put forth the strong similarity between the original Feynman ratchet and a mesoscopic thermoelectric generator with a single conducting channel. 
\end{abstract}

\maketitle
\section{Introduction}
In his reflections on the motive power of heat, Carnot established that heat-to-work conversion is a process that occurs only with a limited efficiency \cite{Carnot1824}. More than a century later, Feynman, in his \textit{Lectures on Physics} \cite{Feynman1963}, thought it was worthwhile to provide a physically transparent explanation of this limitation, basing his analysis on elementary mechanical arguments. The simplest mechanical device that could serve this purpose was that which allows a shaft to rotate only one way to transmit motion: the ratchet and pawl system, originally designed by Smoluchowski (among other demons in the thermodynamic bestiary \cite{Smoluchowski1912}) and known nowadays as a Brownian or Feynman ratchet. The system is composed of an axle attached on one side to vanes immersed in a gas at temperature $T_1$ while the other end is attached to a ratchet wheel with asymmetric teeth immersed in a thermal reservoir at temperature $T_2$. The motion of the wheel is constrained by the presence of a pawl which allows its rotation only in one direction. Further, the axle is also provided with a drum on which is attached a load to lift; this latter exerts a torque on the axle.

According to Feynman \cite{Feynman1963}, if the system is sufficiently small, the Brownian motion of the particles in the reservoir 1 communicates a fluctuating motion to the vanes which in turn provides, through the axle, random rotations of the ratchet wheel. When one of the rotations caused by the fluctuations is sufficiently important to make the pawl jump from one tooth to the next, one observes a breaking of the rotational symmetry: Thermal fluctuations then have a non-vanishing average effect. The axle thus experiences an effective motion of rotation in a given direction (determined by the orientation of the teeth) that might be transferred to the load; thereby, work is produced. As such, Feynman's ratchet may be viewed as one of the possible realizations of Maxwell's demon since its purpose is to extract useful work from the thermal fluctuations in a thermal reservoir. But extracting work from a single thermal reservoir is in contradiction with the second law of thermodynamics: ``\emph{No process is possible whose sole result is the complete conversion of heat into work}'', according to Kelvin statement \cite{Blundell2010}.

To overcome this contradiction, Feynman focused on the rectification process caused by the pawl, which indeed is also affected by thermal fluctuations since the mechanical system is assumed to be sufficiently small to operate on Brownian motion. The temperature fluctuations in the thermal reservoir 2 affect the spring attached to the pawl, necessary to replace this latter on the ratchet wheel after the jump from one tooth to the next. If these fluctuations become important, the pawl may disengage from the ratchet and allow, and even force, the wheel to turn in the direction opposite to that which permits work production. Feynman thus introduced the \emph{necessity} to consider two reservoirs and the exchange of heat between them: work can be extracted only if some heat flows from the hotter to the colder reservoir, satisfying thereby the second law. With his toy model, Feynman obtained much insight into the operation of a heat engine, and could discuss the notions of reversibility and irreversibility, as well as order and entropy. 

Since its publication, Feynman's analysis of the ratchet and pawl has inspired numerous works based on more sophisticated model systems (see, e.g., the studies of Sakaguchi \cite{Sakaguchi1998} and Van den Broeck et al. \cite{VdB2004}, and works cited therein). Feynman's ratchet also serves as a reference model for studies of molecular motors \cite{Astumian1997, Reimann2002, Parrondo2002, Hanggi2009}. Therefore, a finer understanding of this particular heat engine not only is of theoretical interest but may also be useful in the field of biophysics, especially for the determination of the performance of molecular motors. Now, only a handful of studies \cite{Parrondo1996,Magnasco1998,Jarzynski1999} concern the original mechanical device proposed by Feynman. In this article, we revisit the original ratchet and pawl system in light of thermoelectric transport theory. Thermoelectricity has long been recognized as a touchstone for theories of irreversible thermodynamics \cite{deGroot} since it provides sufficiently simple arguments and formalism to rigorously tackle outstanding problems in this field, as shown yet again recently \cite{Apertet2012a}. One purpose of the present work is to show that there is a strong analogy between the properties of Feynman's ratchet and those of a model thermoelectric generator, which can be fruitfully exploited to model autonomous heat engines.

The article is organized as follows. In Sec.~\ref{FR}, we recall the principal features of Feynman's ratchet by giving the constitutive relations of the model. Then, focusing first on the linear approximation of these relations as done by Velasco and co-workers in Ref.~\cite{Velasco2001}, we demonstrate in Sec.~\ref{LD} how the formalism used in thermoelectricity may be adapted to analyze Feynman's ratchet. In particular we introduce the notion of \emph{entropy per tooth} in analogy with the \emph{entropy per carrier} and we discuss the validity of the strong-coupling assumption already questioned in Ref.~\cite{Parrondo1996}. Next, in Sec.~\ref{NLD}, we extend our analysis to the original nonlinear case considered by Feynman, putting forth the distinction between the dynamical response of the system and its ability to dissipate energy. We end the present work by stressing the similitude between Feynman's ratchet and a mesoscopic thermoelectric generator with a single conducting channel.

\section{\label{FR} Constitutive relations}

\begin{figure}
	\centering
		\includegraphics*[width=0.48\textwidth]{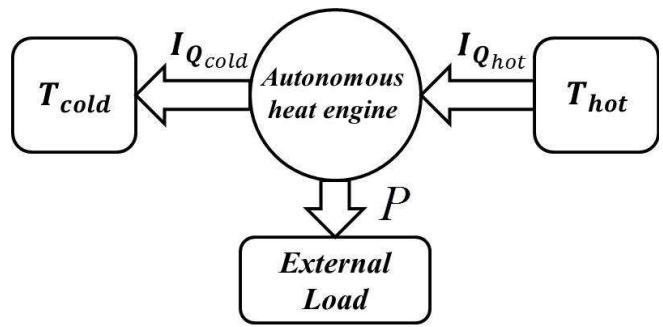}
	\caption{Thermodynamic picture of an autonomous heat engine.}
	\label{fig:figure2}
\end{figure}
A generic autonomous heat engine in contact with two thermal reservoirs is depicted on Fig.~\ref{fig:figure2}. A load, which receives power extracted from the heat flux, embodies the time-independent boundary conditions that drive the operation of the autonomous engine maintained in a nonequilibrium steady state. Onsager's formalism \cite{Onsager1931} is very well suited to analyze such a situation. 

Just as in Ref.~\cite{Feynman1963}, we consider that the reservoir 1 is hotter than the reservoir 2. For clarity, we set $T_1 = T_{\rm hot}$ and $T_2 = T_{\rm cold}$ with $T_{\rm hot} > T_{\rm cold}$. To determine the heat fluxes, we express first the \emph{effective jump frequency} $\dot{N}_{\rm eff}$ from one tooth to the next as a function of the various characteristics, internal and external, of the system. The quantity $\dot{N}_{\rm eff}$ represents the difference between the forward jump frequency $\dot{N}^+$ associated with a lift of the load, and the backward jump frequency $\dot{N}^-$ associated with a fall of the load. A forward jump is obtained when the thermal energy provided to the vanes in the reservoir 1 allows lifting the load but also compressing the spring to let the pawl reach the next tooth. The required energy to compress the spring is denoted $\xi$; the potential energy delivered to the load is the product of the torque $L$ exerted by this load on the axle by the angle $\theta$ between two successive teeth of the wheel. The probability of a forward jump is then supposed to be proportional to $\exp\left[-(\xi+L\theta)/(k_{\rm B} T_{\rm hot})\right]$, where $k_{\rm B}$ is the Boltzmann constant. In the case of a backward jump, only the compression of the spring is involved: The probability of this event is proportional to $\exp\left[-\xi/(k_{\rm B} T_{\rm cold})\right]$. Hence, the jump frequencies are given by
 
\begin{equation}\label{eq:frqsaut}
\begin{array}{l}
\dot{N}^+=\frac{\displaystyle 1}{\displaystyle t}\exp\left(-\frac{\displaystyle \xi+L\theta}{\displaystyle k_{\rm B} T_{\rm hot}}\right)
\\
\dot{N}^-=\frac{\displaystyle 1}{\displaystyle t}\exp\left(-\frac{\displaystyle \xi}{\displaystyle k_{\rm B} T_{\rm cold}}\right)
\end{array}
\end{equation}

\noindent so that the effective jump frequency $\dot{N}_{\rm eff} = \dot{N}^+ - \dot{N}^-$ reads

\begin{equation}\label{eq2}
\dot{N}_{\rm eff} =\frac{1}{t}\left[\exp\left(-\frac{\xi+L\theta}{k_{\rm B} T_{\rm hot}}\right) - \exp\left(-\frac{\xi}{k_{\rm B} T_{\rm cold}}\right)\right]
\end{equation}

\noindent where $t$ is a characteristic time of the system, which is the same in both expressions \eqref{eq:frqsaut}, since one should recover the situation of thermal equilibrium in the absence of the load ($L\theta = 0$), and with a vanishing average rotation of the axle ($\dot{N}^+ = \dot{N}^-$), when both heat reservoirs are at the same temperature ($T_{\rm hot} = T_{\rm cold}$). 

The mean heat fluxes between the device and the thermal reservoirs can be obtained as the product of the effective jump frequency and of the energy extracted (delivered) from (to) the reservoir. The convention used for the orientations of the heat fluxes is displayed on Fig.~\ref{fig:figure2}. Since for each forward jump an energy $\xi+L\theta$ is taken from the hot reservoir while an energy $\xi$ is delivered to the cold reservoir, the heat fluxes are given by
\begin{equation}\label{hflx}
\begin{array}{l}
I_{Q_{\rm hot}} = \left(L\theta + \xi\right) \dot{N}_{\rm eff}
\\
I_{Q_{\rm cold}} = \xi \dot{N}_{\rm eff}
\end{array}
\end{equation}

\noindent Thereby, the power $P$ transferred to the load is

\begin{equation}
P = I_{Q_{\rm hot}} - I_{Q_{\rm cold}} = L\theta \dot{N}_{\rm eff} .
\end{equation}

\noindent Note that this power may be viewed as the product of a generalized flux $\dot{N}_{\rm eff}$ and a generalized thermodynamic force applied to the load $L\theta$.

\section{\label{LD} Linear description}
Because of the exponential dependence of the effective jump frequency $\dot{N}_{\rm eff}$ on the external control parameter $L\theta$, it is difficult to obtain a clear physical picture of the phenomena at stake in the operation of Feynman's ratchet. In order to simplify the problem and hence clarify its analysis, we assume in this section that the energies associated with each jump, namely, $\xi$ and $\xi + L\theta$, remain negligible compared to the thermal energies of the reservoirs $k_{\rm B} T_{\rm hot}$ and $k_{\rm B} T_{\rm cold}$. 

\subsection{Linearization of the jump frequencies}

Using this simplifying hypothesis, the number of forward and backward jumps per unit of time are respectively given by

\begin{equation}
\begin{array}{c}
\dot{N}^+=\frac{\displaystyle 1}{\displaystyle t}\left( 1- \frac{\displaystyle \xi + L\theta}{\displaystyle k_{\rm B} T_{\rm hot}}\right),
\\
\dot{N}^-=\frac{\displaystyle 1}{\displaystyle t}\left( 1- \frac{\displaystyle \xi }{\displaystyle k_{\rm B} T_{\rm cold}}\right),
\end{array}
\end{equation}

\noindent which yield a simplified expression for the effective jump frequency:

\begin{equation}\label{Nefflineaire}
\dot{N}_{\rm eff} = \frac{1}{t}\left( \frac{\xi }{k_{\rm B} T_{\rm cold}}- \frac{\xi + L\theta}{k_{\rm B} T_{\rm hot}}\right) = \frac{L_0\theta - L\theta}{t k_{B} T_{\rm hot}},
\end{equation}

\noindent where, following Velasco and coworkers \cite{Velasco2001}, we introduced the parameter $L_0\theta = \Delta T \xi/T_{\rm cold}$. We see that $L_0\theta$ is the value of the control parameter, through the choice of the load, for which $\dot{N}_{\rm eff}$ vanishes. In the following, $L_0\theta$ will thus be called the \emph{stopping force}.

\subsection{Analogy with thermoelectricity}

The heat fluxes \eqref{hflx} may now be rewritten as

\begin{eqnarray}\label{eq:thermalfluxFR}
I_{Q_{\rm hot}} & = & \left(\frac{\xi}{T_{\rm cold}}\right)T_{\rm hot}\dot{N}_{\rm eff} - tk_{\rm B} T_{\rm hot}\dot{N}_{\rm eff}^2 , \nonumber \\
\\
I_{Q_{\rm cold}} & = & \left(\frac{\xi}{T_{\rm cold}}\right)T_{\rm cold}\dot{N}_{\rm eff} . \nonumber
\end{eqnarray}

\noindent With this form, we notice a strong similitude with the heat fluxes between a thermoelectric engine and the thermal reservoirs with which it is in contact. We do not use here the classical expressions of Ioffe \cite{Ioffe1957}; rather we use the generalized expressions derived in Ref.~\cite{Apertet2013}:

\begin{eqnarray}\label{fluxthermiqueTE}
I_{Q_{\rm hot}} & = & \alpha T_{\rm hot} I + K \Delta T - \gamma R I^2, \nonumber \\
\\
I_{Q_{\rm cold}} & = & \alpha T_{\rm cold} I + K \Delta T +(1-\gamma)R I^2, \nonumber
\end{eqnarray}

\noindent with $I$ being the electrical current, $\alpha$ being the Seebeck coefficient, $K$ being the thermal conductance of the engine, $R$ being its electrical resistance and $\gamma$ being the coefficient of partition of the dissipated heat between the two thermal reservoirs. The electrical current $I$ for a thermoelectric system and the effective jump frequency $\dot{N}_{\rm eff}$ for Feynman's ratchet play similar roles: both are a ``\emph{useful flux}'' in the sense that they are associated with work production. Now, identifying Eqs.~(\ref{eq:thermalfluxFR}) and (\ref{fluxthermiqueTE}) term by term, it is thus possible to find a thermoelectric equivalent to each parameter of the Feynman ratchet. 

\subsubsection{Entropy per tooth}
Various mechanisms such as radiation, conduction, and convection contribute to heat transfer. In thermoelectricity, convection may be associated with the net \emph{useful} flux of electric charges since this mechanism underlies the thermoelectric conversion process \cite{Thomson1856, Apertet2012}: The power delivered by a thermoelectric generator to a load results from the difference between the incoming and outgoing convective heat fluxes across the generator. Now, assuming that the heat transported by convection is proportional to the local temperature $T$, the energy conversion is then only related to the temperature change along the device and one may define a constant quantity $\alpha$, the Seebeck coefficient, such that the heat flux reads  $I_{Q} = \alpha T I$. 

We may adapt the same reasoning to Feynman's ratchet and, with Eq.~\eqref{eq:thermalfluxFR}, introduce an \emph{entropy per tooth}, $\alpha_{\rm FR}$, in analogy with the Seebeck coefficient, sometimes defined as the \emph{entropy per particle}:

\begin{equation}\label{alphaFR1}
\alpha_{\rm FR} = \frac{\xi}{T_{\rm cold}}.
\end{equation}

\noindent This expression depends only on the compression energy $\xi$ of the spring and on $T_{\rm cold}$. It is important to notice that the load has no influence whatsoever on the value of $\alpha_{\rm FR}$. The entropy per tooth $\alpha_{\rm FR}$ could also be obtained by reasoning from the stopping force $L_0\theta$. In the thermoelectric case, this stopping force is given by $\alpha \Delta T$. For Feynman's ratchet, we also find:

\begin{equation}\label{alphaFR2}
L_0\theta = \alpha_{\rm FR} \Delta T
\end{equation}

Now, the fact that both ways yield the same expression for $\alpha_{\rm FR}$ is related to a fundamental property of the system: Eq.~(\ref{alphaFR1}) is based on the analysis of the quantity of heat associated with each jump, and hence it is tightly related to the analog of the Peltier coefficient $\Pi_{\rm FR}$, while Eq.~(\ref{alphaFR2}) is obtained from the response of the system to an applied temperature difference. Since $\alpha_{\rm FR}$ is identical for these two aspects of the energy conversion, we may state that:

\begin{equation}\label{KelvinFR}
\Pi_{\rm FR} = \alpha_{\rm FR} T
\end{equation}

\noindent which means that \emph{Kelvin's second relation also holds for the Feynman's ratchet}.

\subsubsection{Resistance and dissipation}
The comparison of the dissipation terms in Eqs.~\eqref{eq:thermalfluxFR} and \eqref{fluxthermiqueTE} shows that the dissipated heat appears only on the hot side of Feynman's ratchet, which implies that the partition parameter $\gamma$ is 1 for this system. This situation is different from the thermoelectric generator for which $\gamma = 1/2$. Further, for Feynman's ratchet, the dissipation term is proportional to the square of the useful flux $\dot{N}_{\rm eff}^2$, just as Joule heating is proportional to $I^2$. We may thus express the equivalent \emph{dissipative resistance} for Feynman's ratchet as

\begin{equation}
R_{\rm dis}^{\ell} =  t k_{\rm B} T_{\rm hot}.
\end{equation}

\noindent As for $\alpha_{\rm FR}$, this resistance does not depend on the load.

In analogy with the thermoelectric generator for which $I=(\alpha\Delta T - \Delta V)/R$, the relation between the useful flux $\dot{N}_{\rm eff}$ and the generalized force $(L_0\theta - L\theta)$ may take the form

\begin{equation}
\dot{N}_{\rm eff} = (L_0\theta - L\theta) / R_{\rm dyn}^{\ell},
\end{equation}

\noindent where $R_{\rm dyn}^{\ell}$ is a \emph{dynamical resistance} characterizing the proportionality between these two quantities. With Eq.~\eqref{Nefflineaire}, we see that

\begin{equation}
R_{\rm dis}^{\ell} = R_{\rm dyn}^{\ell},
\end{equation}

\noindent which constitutes the simplest formulation of the fluctuation-dissipation theorem. From this simple case, we recover the fact that if a system is linear, the useful flux is proportional to the applied force and that it dissipates power quadratically with the flux as stressed in Ref.~\cite{Callen1951}. We insist here on the fact that these quadratic terms are fully part of the linear description of the engine and may not, in principle, be justified by the introduction of \emph{ad-hoc} nonlinear terms as was done for example in Ref.~\cite{Izumida2012}. As a matter of fact, the so-called minimally nonlinear irreversible thermodynamic models \cite{Izumida2012,Izumida2014} are \emph{ad-hoc} theoretical constructs which find a use when passing from the local to the global scale seems infeasible in a reasonable fashion for some model systems.

For Feynman's ratchet, dissipation is due to the damping mechanism necessary to stop the bouncing of the pawl \cite{Feynman1963}. From a Maxwell's demon viewpoint, one would say that the pawl needs to ``forget'' that it was moved. It is thus surprising that the dissipation occurs in the cold reservoir, where the pawl lies, while on the macroscopic scale it appears that all the dissipated heat is released in the hot reservoir ($\gamma = 1$). 

\subsubsection{Thermal conductance}
Since in Feynman's description \cite{Feynman1963} the axle is assumed to be a perfect thermal insulator, no term related to heat conduction, i.e., heat transfer when $\dot{N}_{\rm eff} = 0$, appears in Eq.~(\ref{eq:thermalfluxFR}): The ratchet thus works under the so-called \emph{strong coupling condition} \cite{Kedem1965}. While Gomez-Marin and Sancho arrive at a similar conclusion in Ref.~\cite{GomezMarin2006}, this result is in contradiction with the article of Parrondo and Espa\~nol who demonstrate that there \emph{are} heat leaks in this engine \cite{Parrondo1996}. But, since they considered a system in which the ratchet wheel and pawl part is replaced by other vanes, which is not the genuine Feynman ratchet configuration, the behavior of their system is obviously quite different: the presence of a temperature difference between the two reservoirs and the choice of a load condition $L\theta = 0$ (which is equivalent to a short-circuit in thermoelectricity), implies that the effective jump frequency is finite. As a result, a convective heat flux is generated. We believe that this convective flux has been erroneously interpreted as a conductive heat transfer, leading Parrondo and Espa\~nol to challenge the idea that Feynman's ratchet may operate in the strong coupling regime. As the effective jump frequency vanishes for $L\theta = L_0\theta$ the heat exchange between the reservoir also vanishes. For this particular working condition, in the absence of heat leaks, the Carnot efficiency may be reached as for a thermoelectric generator operating under the strong coupling condition \cite{Gordon1991}. Note that, just as Feynman did, we neglect here the contribution of heat transfer due to the difference of the pawl trajectories along a single tooth for backward and forward rotation, which indeed may be another source of heat leakage \cite{Magnasco1998}.

Although we have just demonstrated from the constitutive relations that there are no heat leaks, the strong coupling condition for Feynman's ratchet still has to be questioned. Indeed there is a discrepancy between the constitutive relations and the global description of the system. As a matter of fact, the system's properties are such that in reservoir 1 low energy particles do not contribute to the energy exchange; but this is not the case for the high energy particles. According to the mathematical description given by Feynman, only particles with an energy exactly equal to $L\theta + \xi$ can contribute to the energy exchange. Since this theoretical fact has no physical justification, the possibility of actually achieving the strong-coupling condition must be checked rigorously, especially when high-energy particles are involved: the thermal conductance under vanishing effective jump frequency might then be no longer negligible.

\subsection{Discussion of the model}
Equation~\eqref{fluxthermiqueTE} gives the heat fluxes at the global scale of the considered thermoelectric generator model; it was derived in Ref.~\cite{Apertet2013}, where we made the assumption of local equilibrium to use the force-flux formalism \cite{Onsager1931} combined with the heat equation to satisfy the principle of conservation of the energy. As Callen provided the correspondence between the kinetic coefficients used by Onsager and the more useful thermoelectric parameters $\alpha$, $R$ and $K$ \cite{Callen1948}, Eq.~(\ref{fluxthermiqueTE}) is an Onsager-Callen description of a thermogenerator. Now, our analysis of Feynman's ratchet presented above is based on an analogy with the Onsager-Callen formalism, and we must discuss the relevance in this context of some of the hypotheses that apply in thermoelectricity.

Feynman's ratchet has already been studied with Onsager's formalism by Gomez-Marin and Sancho \cite{GomezMarin2006}. Even if their analysis is rather based on Sakaguchi's model of the ratchet \cite{Sakaguchi1998}, it also applies to the original Feynman ratchet. In their article, Gomez-Marin and Sancho express the kinetic coefficients $L_{ij}$ of the system considering that the thermodynamical forces are $X_1 = F/T_{\rm hot}$ and $X_2 = \Delta T/T_{\rm hot}$, where $F$ has a role similar to that of $L_0\theta - L\theta$. We see that these expressions do not correspond to local definitions since the temperature of the system is supposed to be always $T_{\rm hot}$. This example illustrates the fact that, as discussed in Ref.~\cite{Apertet2013}, it is impossible to rigorously study an energy-conversion process based on coupled fluxes with a straight Onsager approach. Indeed, within this framework, one must focus on the local scale but not on the system as a whole. However, it is possible to move from the local to the global scale and obtain Eq.~(\ref{fluxthermiqueTE}) as long as the principle of energy conservation is rigorously satisfied.

To proceed with the discussion it is useful to recall that the hypothesis of microreversibility was used to prove the reciprocal relations at the heart of Onsager's theory of irreversible thermodynamics \cite{Onsager1931,Onsager1931b}, and that the same hypothesis was also used later to merge into one formalism known as the Onsager-Machlup theory \cite{OnsagerMachlup}, the phenomenology of irreversible processes and the stochastic analysis of the spontaneous fluctuations of thermodynamic variables. The microreversibility hypothesis applies correctly to a thermodynamic system undergoing irreversible changes, only at the local scale, which implies that the said system is in a situation of near equilibrium, where Onsager's force-flux formalism yields a satisfactory account of the physical processes at work. The description of a thermodynamic system not too far from equilibrium may thus be done with the assumption of local equilibrium ~\cite{Pottier2007}. More precisely, the system is viewed as an ensemble of regions or cells of intermediate sizes, i.e. large enough to be considered as thermodynamic subsystems with variables thus defined \emph{locally}, and in contact with their environment (the cells are open to energy and matter transport), but sufficiently small to ensure little variation of the local thermodynamic variables.

In the case of Feynman's ratchet we derived constitutive relations similar to those obtained for a thermoelectric systems but there is no such thing as a local equilibrium for this system. Our analysis is based on an analogy between two different systems, and it may seem rather surprising to obtain the results we showed above without the assumption of local equilibrium for Feynman's ratchet (in particular no thermodynamic variable has been associated with the axle), while this assumption is crucial for a thermoelectric generator. We thus conclude that, despite the fact it was first obtained with an Onsager-Callen approach, Eq.~\eqref{fluxthermiqueTE} has a much broader scope than that of linear force-flux formalism since the assumption of local equilibrium of the system is superfluous.

As regards the analog to the second Kelvin relation, Eq.~\eqref{KelvinFR}, one may wonder what microscopic reversibility, which is an essential assumption in the case of thermoelectricity, means for the Feynman's ratchet. Our view is that the condition of microscopic reversibility might be linked to the necessity for the characteristic times $t$ in Eq.~\eqref{eq:frqsaut}, to be identical. The justification for this identity originates in the condition of thermal equilibrium with $\Delta T = 0$ and $L\theta = 0$; we thus consider that the second Kelvin relation in this case stems from the zeroth law of thermodynamics.

The angle $\theta$ is an internal parameter of the system, which appears in the definition of the generalized force $L\theta$ associated with the load. This definition simplifies the physical description of the system and has no impact on our derivations. Actually $\theta$ plays the same role for Feynman's ratchet as that of the elementary electrical charge $e$ in thermoelectricity: For convenience, and without any consequence, the electrical current $I = - eI_{\rm particle}$ has been considered instead of the particle current $I_{\rm particle}$ and the potential difference $\Delta V$ has been considered instead of the electrochemical potential difference $\Delta \mu = - e\Delta V$.

Finally, we stress that a generalized configuration of the Feynman ratchet, which accounts for dissipative thermal contacts between the engine and the heat reservoirs, must be analyzed with care, as in this case the temperatures $T_1$ and $T_2$ are no longer equal to $T_{\rm hot}$ and $T_{\rm cold}$ and depend on the working conditions, so $\alpha_{FR}$ and $R_{\rm dis}^{\ell}$ do: The linear behavior is thus lost.

\subsection{Efficiency at maximum power}

The efficiency at maximum power for the linear model of Feynman's ratchet was derived by Velasco and co-workers \cite{Velasco2001}, and they named it the Feynman efficiency:

\begin{equation}
\eta_{\rm Feynman} = \frac{\eta_{\rm C}}{2-\eta_{\rm C}}
\end{equation}
\noindent where $\eta_{\rm C}$ is the Carnot efficiency. This result may easily be recovered using the general expression initially derived in Ref.~\cite{Schmiedl2008}:
\begin{equation}
\eta_{\rm SS} = \frac{\eta_{\rm C}}{2 - \gamma \eta_{\rm C}}
\end{equation} 
\noindent since we have just found that the partition coefficient of the dissipated heat $\gamma$ is 1 for the Feynman's ratchet, which is the most favorable configuration for maximizing the efficiency at maximum power.

\section{\label{NLD} Non-linear model}

We turn now to the nonlinear properties of Feynman's ratchet. Although we necessarily move beyond the linear regime studied above, we adopt essentially the same approach to retain the clarity of our analysis so far, basing our reasoning on analogies with thermoelectricity.

\subsection{Modified relationship between $R_{\rm dis}$ and $R_{\rm dyn}$}

The effective jump frequency, Eq.~\eqref{eq2}, takes the following form

\begin{equation}\label{eq:freqnonlineaire}
\dot{N}_{\rm eff} =t^{-1}\exp\left(-\frac{\xi+L\theta}{k_{\rm B} T_{\rm hot}}\right)\left[1 - \exp\left({-\frac{L_0\theta-L\theta}{k_{\rm B} T_{\rm hot}}}\right)\right]
\end{equation}

\noindent using the notations introduced in the previous section. It vanishes for $L\theta = L_0\theta$ as for the linear model; so the quantity $L_0\theta$ is still viewed as the stopping force in the nonlinear model. The heat fluxes are given by

\begin{equation}
\begin{array}{l}
I_{Q_{\rm hot}} = \alpha_{\rm FR} T_{\rm hot}\dot{N}_{\rm eff} - \left[L_0\theta - L\theta\right]\dot{N}_{\rm eff}  \\
\\
I_{Q_{\rm cold}}= \alpha_{\rm FR} T_{\rm cold}\dot{N}_{\rm eff}  \\
\end{array}
\end{equation}

To keep the same form for the heat flux as in Eq.~(\ref{eq:thermalfluxFR}), we give the following general definition of the resistance associated with the dissipation process:

\begin{equation}\label{Rdissip}
R_{\rm dis} = \frac{L_0\theta - L\theta}{\dot{N}_{\rm eff}}.
\end{equation}

\noindent One may recover $R_{\rm dis}^{\ell}$ from this expression assuming a linear regime.

We also give a different definition of the dynamical resistance $R_{\rm dyn}$, which was interpreted above as the link between the generalized force and the useful flux. In the nonlinear model, we do not consider the global behavior of the system but rather the vicinity of the considered working point: $R_{\rm dyn}$ is the derivative of the generalized force $L_0\theta - L\theta$ with respect to the useful flux $\dot{N}_{\rm eff}$:

\begin{equation}
R_{\rm dyn}= - \frac{{\rm d}(L_0\theta - L\theta)}{{\rm d}\dot{N}_{\rm eff}} = - \frac{{\rm d}(L\theta)}{{\rm d}\dot{N}_{\rm eff}}.
\end{equation}

\noindent For Feynman's ratchet the two resistances are

\begin{equation}
R_{\rm dyn}= t k_{\rm B} T_{\rm hot} \exp\left(\frac{\xi + L\theta}{k_{\rm B} T_{\rm hot}}\right)
\end{equation}

\noindent and 

\begin{equation}\label{eq:RdissipNL}
R_{\rm dis}= \frac{L_0\theta-L\theta}{k_{\rm B} T_{\rm hot}\left[1 - \exp\left(-\frac{L_0\theta-L\theta}{k_{\rm B} T_{\rm hot}}\right)\right]}~ R_{\rm dyn}.
\end{equation}

\begin{figure}
	\centering
		\includegraphics*[width=0.48\textwidth]{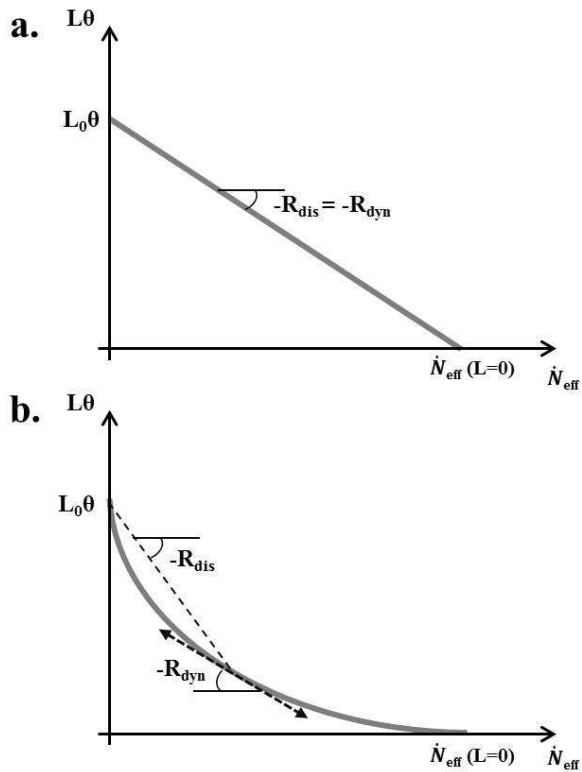}
	\caption{Behavior of the linear (a) and nonlinear (b) Feynman's ratchet. The working point is selected by setting the value of the torque $L$ defined by the load.}
	\label{fig:figure3}
\end{figure}

\noindent In contrast to their linear counterparts, both resistances depend on $L\theta$ and hence on the working conditions. Further, they are not equal except in the quasistatic limit, reached for $L\theta = L_0\theta$,  where $\dot{N}_{\rm eff}$ vanishes. For this particular engine, we notice that $R_{\rm dis} \geq R_{\rm dyn}$. The main differences between the linear and nonlinear models of Feynman's ratchet are summarized on Fig.~\ref{fig:figure3}. 

\subsection{Analogy with a mesoscopic thermoelectric generator}

\begin{figure}
	\centering
		\includegraphics*[width=0.40\textwidth]{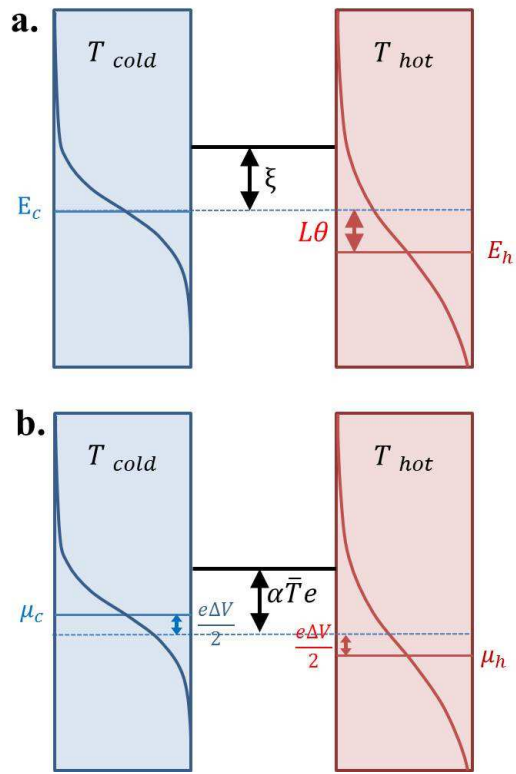}
	\caption{Comparison of Feynman's ratchet (a) and a thermoelectric generator with a single conducting channel (b) using mesoscopic physics formalism.}
	\label{fig:figure4}
\end{figure}

The analogy with a thermoelectric generator may be carried out further, even in the nonlinear case. To this purpose, we consider a mesoscopic system made of a single ballistic conducting channel connecting two reservoirs at different temperatures such as that described by Humphrey and co-workers in Refs.~\cite{Humphrey2002,Humphrey2005}. If the conducting channel is at an energy sufficiently above the electrochemical potentials, i.e., the Fermi levels, of the reservoirs, then the electron distribution may be approximated by a Boltzmann distribution which is similar to the probability law for a jump from one tooth to the next in Feynman's ratchet. As stressed in the previous section, only particles with a very specific energy may contribute to the conversion process; then it possible to model Feynman's ratchet like a mesoscopic device as illustrated in Fig.~\ref{fig:figure4}(a). Comparison with a genuine mesoscopic thermoelectric generator, in Fig.~\ref{fig:figure4}(b), shows that the only difference between these two systems is the choice of the energy reference. In the case of Feynman's ratchet the energy level of the cold reservoir is chosen as a reference, and it is not modified when the load is changed; whereas in the case of the thermoelectric generator the energy reference is set as the mean of the Fermi levels in the cold and the hot reservoirs, $\mu_{\rm cold}$ and $\mu_{\rm hot}$: When the load is modified, the potential difference $\Delta V$, and both $\mu_{\rm cold}$ and $\mu_{\rm hot}$ are modified. This discrepancy between the two models explains the difference for the value of the partition coefficient of the dissipated heat $\gamma$, whose value is $1$ for Feynman's ratchet, but $1/2$ for the thermoelectric generator. In both cases the ``conducting channel'' is placed above the energy reference at a constant energy value: $\xi$ for Feynman's ratchet and $\alpha \overline{T} e$ for the thermoelectric generator \cite{Mahan1996}, with $\overline{T} = (T_{\rm hot}+T_{\rm cold})/2$.

From this comparison, we may also propose a different viewpoint on the characteristic time $t$. Indeed, $1/t$ appears to play the same role as the transmission coefficient of the channel. It is thus easy to understand why this value is the same for both reservoirs in Eq.~\eqref{eq:frqsaut}: it is an intrinsic property of the \emph{link between the reservoirs} rather than a property of each reservoir. We end this section by noticing that, as a matter of fact, Feynman was probably the first to use a result that was derived later by Mahan and Sofo \cite{Mahan1996} and discussed in Refs.~\cite{Humphrey2002,Humphrey2005}: a single ballistic channel, corresponding to a Dirac distribution for the transmission function, is a way to reach the strong-coupling condition, i.e., $ZT \longrightarrow \infty$ for a thermoelectric system.\\

\section{\label{Conclu} Summary and concluding remarks}
We have shown how to adapt the formalism of Onsager andCallen in thermoelectric transport theory to analyze the linear model of Feynman's ratchet. We defined an \emph{entropy per tooth} in analogy with the entropy per particle. We generalized and extended our analysis beyond the linear model, stressing the necessary distinction between the dynamical resistance $R_{\rm dyn}$ and the dissipation resistance $R_{\rm dis}$. The present analysis of Feynman's ratchet should prove useful for the modeling of autonomous heat engines. Indeed our work puts forth the need for a transition from Onsager's formalism, for which the assumption of local equilibrium is mandatory, to a more general description where the global behavior of the engine is considered, and thus is closer to Callen's description of the thermoelectric phenomena \cite{Callen1948}. The Feynman ratchet example illustrates the facts that this transition is simple and that the assumption of local equilibrium is not required to derive the macroscopic properties ($\alpha_{FR}$, $R_{\rm dis}$, $R_{\rm dyn}$ and $K$). In principle, thermoelectric equivalent parameters may be derived for various systems; this would allow the direct use of the well established results in thermoelectricity, in particular for finite-time optimization \cite{Gordon1991, Apertet2013}, rather than an elaborate investigation  from scratch for each system. Actually, the thermoelectric analogy has already been implicitly used to describe the coupled transport of heat and particles in the case of cold atoms \cite{Brantut2013}. This approach may also be useful to interpret recent results concerning ion-exchange membranes \cite{Biesheuvel2014}. 

To conclude this article, it is instructive to recall that theoretical studies of Feynman's ratchet, a Brownian system mathematically characterized by a Wiener process, may be tackled with various approaches. It is worth mentioning that those based on the so-called Onsager-Machlup theory \cite{OnsagerMachlup} are particularly well suited to this purpose. The Onsager-Machlup variational approach was developed in a number of works, and an effective action principle for nonequilibrium dynamics was proposed \cite{Eyink1996} and used to show how linear stochastic models may describe nonlinear dynamical systems \cite{Eyink1998}. Hence our analyses of the nonlinear behavior of Feynman's ratchet could be revisited using the effective action principle. Our purpose in this article was to propose a physically transparent analysis of the linear and nonlinear dynamics of Feynman's ratchet as originally described in the \textit{Lectures on Physics} \cite{Feynman1963}; our view being that thermoelectricity provides an interesting approach owing to the analogies we put forth in the various sections of the paper. In a forthcoming presentation, we will build on the nonlinear model of Feynman's ratchet to derive a general expression of the efficiency at maximum power of an autonomous engine \cite{Apertet2014}.


\begin{thebibliography}{10}

\bibitem{Carnot1824} S. Carnot, \emph{R\'eflexions sur la Puissance Motrice du Feu, et sur les Machines Propres \`a D\'evelopper cette Puissance} (Bachelier, Paris, 1824).

\bibitem{Feynman1963} R.P. Feynman, R.B. Leighton, and M. Sands, \emph{The Feynman Lectures on Physics}, (Addison-Wesley, Reading, MA, 1963), Vol. I, Chap. 46.

\bibitem{Smoluchowski1912} M. Smoluchowski, Phys. Z. {\bf 13}, 1069 (1912).

\bibitem{Blundell2010} S. J. Blundell and K. M. Blundell, \emph{Concepts in Thermal Physics}, 2nd ed. (Oxford University Press, New York, 2010). 

\bibitem{Sakaguchi1998} H. Sakaguchi, J. Phys. Soc. Jpn. {\bf 67}, 709 (1998).

\bibitem{VdB2004} C. Van den Broeck, R. Kawai, and P. Meurs, Phys. Rev. Lett. {\bf 93}, 090601 (2004).

\bibitem{Astumian1997} R. D. Astumian, Science {\bf 276}, 917 (1997).

\bibitem{Reimann2002} P. Reimann, Phys. Rep. {\bf 361}, 57 (2002).

\bibitem{Parrondo2002} J. M. R. Parrondo and B. J. de Cisneros, Appl. Phys. A {\bf 75}, 179 (2002).

\bibitem{Hanggi2009} P. Hanggi and F. Marchesoni, Rev. Mod. Phys. {\bf 81}, 387 (2009).

\bibitem{Jarzynski1999} C. Jarzynski and O. Mazonka, Phys. Rev. E {\bf 59}, 6448 (1999).

\bibitem{Magnasco1998} M. O. Magnasco and G. Stolovitzky, J. of Stat. Phys. {\bf 93}, 615 (1998).

\bibitem{Parrondo1996} J. M. R. Parrondo and P. Espa\~nol, Am. J. Phys. {\bf 64}, 1125 (1996).

\bibitem{deGroot} S. R. de Groot, \emph{Thermodynamics of Irreversible Processes} (Intersciences, New York, 1958).

\bibitem{Apertet2012a} Y. Apertet, H. Ouerdane, C. Goupil, and Ph. Lecoeur, Phys. Rev. E {\bf 85}, 031116 (2012).

\bibitem{Velasco2001} S. Velasco, J. M. M. Roco, A. Medina and A Calvo Hern\'andez, J. Phys. D {\bf 34}, 1000 (2001).

\bibitem{Onsager1931} L. Onsager, Phys. Rev. {\bf 37}, 405 (1931).

\bibitem{Ioffe1957} A. F. Ioffe, \emph{Semiconductor Thermoelements and Thermoelectric Cooling} (Infosearch, London, 1957).

\bibitem{Apertet2013} Y. Apertet, H. Ouerdane, C. Goupil, and Ph. Lecoeur, Phys. Rev. E {\bf 88}, 022137 (2013).

\bibitem{Thomson1856} W. Thomson, Philos. Trans. R. Soc. London {\bf 146}, 649 (1856).

\bibitem{Apertet2012} Y. Apertet, H. Ouerdane, C. Goupil, and Ph. Lecoeur, J. Phys.:Conf. Ser. {\bf 395}, 012103 (2012).

\bibitem{Callen1951} H. B. Callen and T. A. Welton, Phys. Rev. {\bf 83}, 34 (1951).

\bibitem{Izumida2012} Y. Izumida and K. Okuda, Europhys. Lett. {\bf 97}, 10004 (2012).

\bibitem{Izumida2014} Y. Izumida, K. Okuda, J. M. M. Roco, and A. Calvo Hern{\'a}ndez, arXiv:1405.6777.

\bibitem{Kedem1965} O. Kedem and S. R. Caplan, Trans. Faraday Soc. {\bf 61}, 1897 (1965).

\bibitem{GomezMarin2006} A. Gomez-Marin and J. M. Sancho, Phys. Rev. E {\bf 74}, 062102 (2006).

\bibitem{Gordon1991} J. M. Gordon, Am. J. Phys. {\bf 59}, 551 (1991).

\bibitem{Callen1948} H. B. Callen, Phys. Rev. {\bf 73}, 1349 (1948).

\bibitem{Onsager1931b} L. Onsager, Phys. Rev. {\bf 38}, 2265 (1931).

\bibitem{OnsagerMachlup} L. Onsager and S. Machlup, Phys. Rev. {\bf 91}, 1505 (1953).

\bibitem{Pottier2007} N. Pottier, \emph{Non-equilibrium Statistical Physics }(Oxford University Press, Oxford, 2010), p. 32.

\bibitem{Schmiedl2008} T. Schmiedl and U. Seifert, Europhys. Lett. {\bf 81}, 20003 (2008).

\bibitem{Mahan1996} G. Mahan and J. Sofo, Proc. Natl. Acad. Sci. U.S.A. {\bf 93}, 7436 (1996).

\bibitem{Humphrey2002} T. E. Humphrey, R. Newbury, R. P. Taylor, and H. Linke, Phys. Rev. Lett. {\bf 89}, 116801 (2002).

\bibitem{Humphrey2005} T. E. Humphrey and H. Linke, Phys. Rev. Lett. {\bf 94}, 096601 (2005).

\bibitem{Brantut2013} J.P. Brantut, C. Grenier, J. Meineke, D. Stadler, S. Krinner, C. Kollath, T. Esslinger, and A. Georges, Science {\bf 342}, 713 (2013).

\bibitem{Biesheuvel2014} P. M. Biesheuvel, D. Brogioli, and H. V. M. Hamelers, arXiv:1402.1448.

\bibitem{Eyink1996} G. L. Eyink, Phys. Rev. E {\bf 54}, 3419 (1996)

\bibitem{Eyink1998} G. L. Eyink, Phys. Rev. E {\bf 58}, 6975 (1998). 

\bibitem{Apertet2014} Y. Apertet, H. Ouerdane, C. Goupil, and Ph. Lecoeur (unpublished); Y. Apertet, Ph.D. thesis, Universit\'e Paris-Sud, 2013 (in French).

\end{thebibliography}
\end{document}